\documentclass[modern]{aastex631}
\shortauthors{C. Salter et al.}
\begin{document}

\title{The discovery and evolution of a radio continuum and excited-OH spectral-line outburst in the nearby galaxy NGC 660}

\author{C.~J. Salter} 
\affiliation{Retired, Arecibo Observatory, HC3 Box 53995, Arecibo, PR 00612, USA}
\author{T. Ghosh} 
\affiliation{Retired, Green Bank Observatory, P.O. Box 2, Green Bank, WV 24944, USA}
\author{R.~F. Minchin}
\affiliation{National Radio Astronomy Observatory, 1003 Lopezville Rd.,
P.O. Box O, Socorro, NM 87801, USA}
\author{E. Momjian}
\affiliation{National Radio Astronomy Observatory, 1003 Lopezville Rd.,
P.O. Box O, Socorro, NM 87801, USA}
\author{B. Catinella}
\affiliation{ICRAR, The University of Western Australia, 35 Stirling
Highway, Crawley, WA 6009, Australia}
\author{M. Lebron}
\affiliation{University of Puerto Rico, P.O. Box 23323 , San Juan, PR 00931-3323, USA}
\author{M.S. Lerner}
\affiliation{Onsala Space Observatory, SE-439 92 Onsala, Sweden}
\correspondingauthor{C.~J. Salter} 
\email{ }
\email{csalter.wfc@gmail.com, tapasig91@gmail.com, rminchin@nrao.edu, emomjian@nrao.edu,}
\email{barbara.catinella@uwa.edu.au, mayra.lebron3@upr.edu, mikael.lerner@chalmers.se}
\begin{abstract}
Arecibo 305-m Telescope observations between 2008 and 2018 detected a radio continuum
and spectral-line outburst in the nearby galaxy, NGC~660. Excited-OH
maser emission/absorption lines near 4.7~GHz, and H$_2$CO absorption at
4.83~GHz varied on time-scales of months. Simultaneously, a continuum
outburst occurred in which a new compact component appeared, with a
GHz-peaked spectrum and a 5-GHz flux density 
that rose to a peak value of about 500 mJy 
from 2008.0 to 2012.0. Follow-up interferometric continuum images from the
Very Large Array (VLA) at ~10~GHz of this new continuum component
determined it to be located at the nucleus of NGC~660. Subsequent High
Sensitivity Array (HSA) line and continuum VLBI observations of the
NGC~660 nucleus revealed a morphology that appears to be consistent with rapidly-precessing, mildly-relativistic
jets from the central black hole. While requiring detailed modeling, this strongly suggests that the outburst is due to nuclear
activity. From its time-scale, the shape of the continuum light-curve,
and the milliarcsec radio structure, the most likely cause of the outburst is 
AGN-type activity of accretion of a gas cloud onto
the central black hole.
\end{abstract}


\section{Introduction}
At a redshift of $z = 0.003$, the ``peculiar" galaxy NGC~660 has  highly unusual properties. While classified as type SBa, its gas-rich inner disk has a composite LINER/H\,{\sc ii}
spectrum, and its far infra-red luminosity of log(L$_{\rm{FIR}})$~=~10.3,
makes it a Luminous Infra-Red Galaxy (LIRG). In addition, it possesses
a component that is strongly warped out of the plane of the main disc,
and an outer quasi-polar ring. NGC~660 also shows strong 
(S$_{\rm{1.4~GHz}}$~=~373~mJy) radio continuum emission of size about $32\arcsec \times
15\arcsec$, and a more compact central double/edge-on  component of
size $\sim~4\arcsec$. With 210-mas resolution, \citet{fil02}
claimed to have located a compact core of 8.4-GHz flux density
$\sim$~3~mJy. The radio continuum spectrum in the centimeter range
is that expected from synchrotron emission, (see the solid black line in Fig. 3.)

In 2008, the current authors detected a surprisingly rich cm-wavelength (1.1 to 10 GHz) molecular-line spectrum from the prototypical Ultra-Luminous Infra-Red Galaxy (ULIRG), Arp~220, \citep{sal08}.  This included the presence of such species as the ``prebiotic molecule'' CH$_{2}$NH, CH and H$_{2}$CO in emission, with transitions of HCN (v$_{2}$=1) and excited-OH, plus probably CH$_{3}$OH, in absorption. 
This was then followed up with Arecibo 4.3 -- 5.3~GHz observations of 20 ``Arp~220-like'' LIRGs. This frequency range contains a rich part of the spectrum of Arp~220, with transitions of HCN (v$_{2}$=1) at 4488~MHz, excited OH $2\Pi_{1/2}$, J=1/2 at 4660, 4750 and 4765~MHz, H$_{2}$CO at 4829~MHz and CH$_{2}$NH at 5290~MHz.  The aim was to determine the occurrence rate of molecular spectra similar to
Arp~220 among (U)LIRGS. 

The sample of 20 galaxies were selected on the basis of having
known $\lambda$18-cm~OH-megamasers \citep{che07}, formaldehyde lines \citep{ara04},
compact starbursts \citep{con91} or far-IR emission lines similar to Arp~220 \citep{arm90}.
The sample was also limited to $z < 0.033$ and $S_{6cm} > 20$~mJy.  NGC~660 was 
included on the basis of having both a formaldehyde absorption feature and a far-IR spectrum similar to Arp~220.  However, it was not noted as having either $\lambda$18-cm~OH-megamaser emission or the extremely high star formation rate of Arp~220.

Two of the sample galaxies were found to possess almost identical spectra to Arp~220 \citep{min09}.
However, the detailed line spectrum of NGC~660 did not resemble that of
Arp~220. Nevertheless,  it did yield the biggest surprise of the project
in the form of excited-OH transitions 
at 4660, 4750 and 4765 MHz showing unprecedented transient behavior in intensity and line profiles such as had never been previously reported for these transitions in extragalactic objects.  In addition, the variability of these OH spectral-lines turned out to be closely mimicked over a ten-year period by secular changes in the  cm-wavelength radio continuum emission. Here, we present decade-long
single-dish and interferometric observations of this enigmatic event.

The observations we present in this paper demonstrate that the event causing the remarkable transient spectral-line and continuum emission seen by us in NGC~660 is occurring at the very center of this galaxy. Continuum variability of many extragalactic redio sources, and in particular blazars,
quasars and other compact radio sources, have been recognized through detailed
monitoring of flux densities at centimetric and millimetric wavelenths, \citep[e.g.] [and refererences therein.] {all17} Such variability is found on all time scales from intra-day to decades. \cite{hov08} studied a large sample of active galactic nuclei (AGNs) across the  cm- and mm-wave window, identifying
distinct transient flares in their flux density curves. They calculated that the duration for 150 flares at 37 GHz and 31 flares at 230 GHz varied from 0.3 to 13.2 yr, noting that the rise and decay times of the flares seem approximately equal, with the decay time typically being 1.3 times the rise time.

Also showing transient continuum radio emission, particularly from within the nuclear regions of LIRGs, are young supernovae \citep[e.g.][]{bie21}, and tidal disruption events (TDEs). The latter occur when a star passes within the tidal radius of the central supermassive black hole (SMBH) of a galaxy,
disrupting the star and ejecting some 50\% of its material, possibly in the form of jets. A recent review of the radio properties of TDEs was published by \cite{ale20}.

In this paper, we present the broad details of our NGC~660  observations and their results. Sections 2 and 3 detail the Arecibo, single-dish spectral-line and radio-continuum monitoring, and present the discovery of the NGC 660 ``Outburst". In Section 4, we discuss new VLA and High Sensitivity Array (HSA)-VLBI images,  with the latter imaging both continuum and spectral-line data. These images help resolve the milliarcsecond structure of the central region of this galaxy. Phenomenological discussion regarding a plausible cause for the  outburst follows in Section 5, and the conclusions are presented in Section 6.

The presentation of a full wide-frequency spectral-line survey of NGC 660, plus further spectral-line and continuum HSA high-resolution images, are deferred to a later paper. Detailed physical modeling
of this source, including consideration of observations at wavelengths other than the radio, is also deferred to  this future publication.

\section{Arecibo Observations and Analysis}

Over a year between December 2007 and 2008, NGC~660 and the 19 other ``Arp 220-like'' galaxies were observed with the 305-m radio telescope at Arecibo Observatory using the WAPP spectrometer
in its dual-board mode, giving $8 \times 100$-MHz wide subbands.  These were placed within a 1-GHz frequency range centered at 4.8~GHz to obtain coverage of a number of spectral lines of interest, including the excited-OH lines at 4660, 4750 and 4765 MHz.  The observations were
made in Double Position Switching \citep[DPS;] [] {gho02} mode, with a continuum calibration (reference) source used to both remove standing waves from the spectrum and calibrate the flux-density scale.  For NGC~660, the reference source selected was a non-variable (compact steep spectrum)
standard radio calibrator B0138+136 (3C~49). NGC~660 and 3C~49 are separated by only 0.5 degrees on the sky, and by only 0.25 degrees in declination. Hence, we expect them to have possessed very similar telecope pointing errors, atmospheric opacities, and telescope spectral baseline ripples 
for all DPS observations. This will have resulted in very accurate relative flux-density calibration from epoch to epoch, which we estimate to be better than 5\%.  We believe that the flux densities that we derive for NGC~660 agree with the absolute scale of \cite{baa77} to better than 10\%.
The data were analyzed using the standard Arecibo AOIDL package written by Phil Perillat.

After the discovery of transient emission in NGC~660 (see Section~\ref{sec3}), follow-up observations
were made in December 2010 and on three occasions in late 2011. Among the observations made on 31~December 2011 were a set that covered 1.1 - 10 GHz using the Arecibo L-, S-low, C-, C-high and X-band receivers in order to obtain a detailed continuum spectrum for the galaxy at that epoch.  
A similar detailed continuum spectrum for NGC~660 
was measured on 14~August 2014. From 2013, NGC~660 was monitored on a semi-regular
basis, both in respect of its C-band spectral lines, and its continuum
flux density up until the end of 2018. However, the 2018 observations, taken after the
Arecibo dish surface was damaged by Hurricane Maria, were of lower quality
and are not included in the present analysis. The results presented here run up
until August 2017. In addition, a full, deep Arecibo spectral scan from 1.1 to 
10 GHz was observed in November and December 2012, results from which will
be presented in a later publication.

\section{Discovery of the Outburst in NGC~660}
\label{sec3}

\begin{figure*}[ht]
\epsscale{0.82}
\plotone{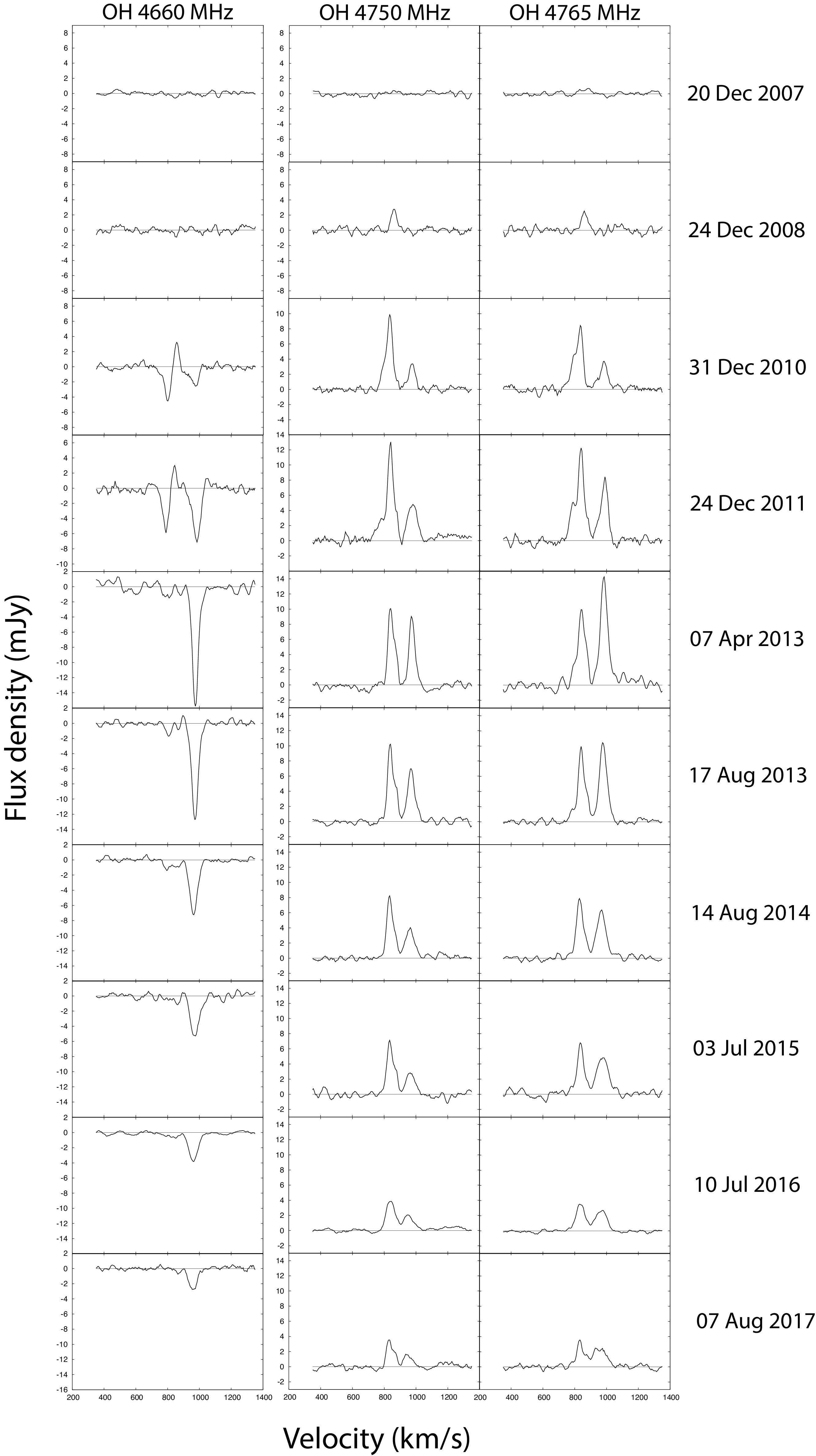}
\caption {Sample excited-OH spectra of NGC~660 from
late 2007 to August 2017. The columns (left to right) show the 4660-,
4750- and 4765-MHz lines respectively, while time increases from top to
bottom. Together, they display large secular changes in these molecular
lines in NGC~660. The heliocentric velocity resolution is $10\,{\rm km}\,{\rm s}^{-1}$.
}
\label{fig1}
\end{figure*}

Spectra of NGC~660 from observing epochs in Dec 2007 and Aug 2008 showed little of
interest apart from an already known H$_{2}$CO absorption line \citep [e.g.] [] {ara04},
and weak hydrogen recombination line emission.
However, the spectrum taken in Dec 2008 revealed a significant
detection of emission from the excited-OH main line at 4750~MHz and its
satellite line at 4765~MHz. Follow-up observations in Dec 2010 not only
confirmed the presence of this emission, but also showed that, (a) the two
emission lines had increased in intensity by almost a factor of four, (b)
additional excited-OH line components, especially at higher velocity,
had appeared for both of these transitions, and (c) counterparts of
these components were also seen in the satellite line at 4660~MHz,
mostly in absorption, but with a central feature in emission (Fig.
\ref{fig1}).

\begin{figure*}[ht]
\hspace{1 cm}
\plottwo{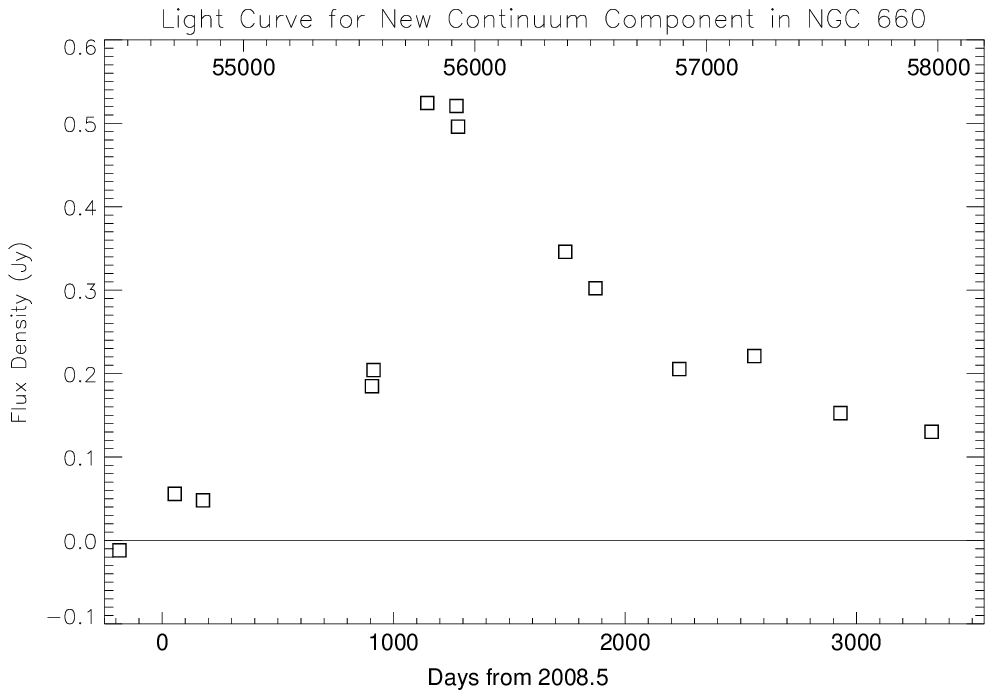}{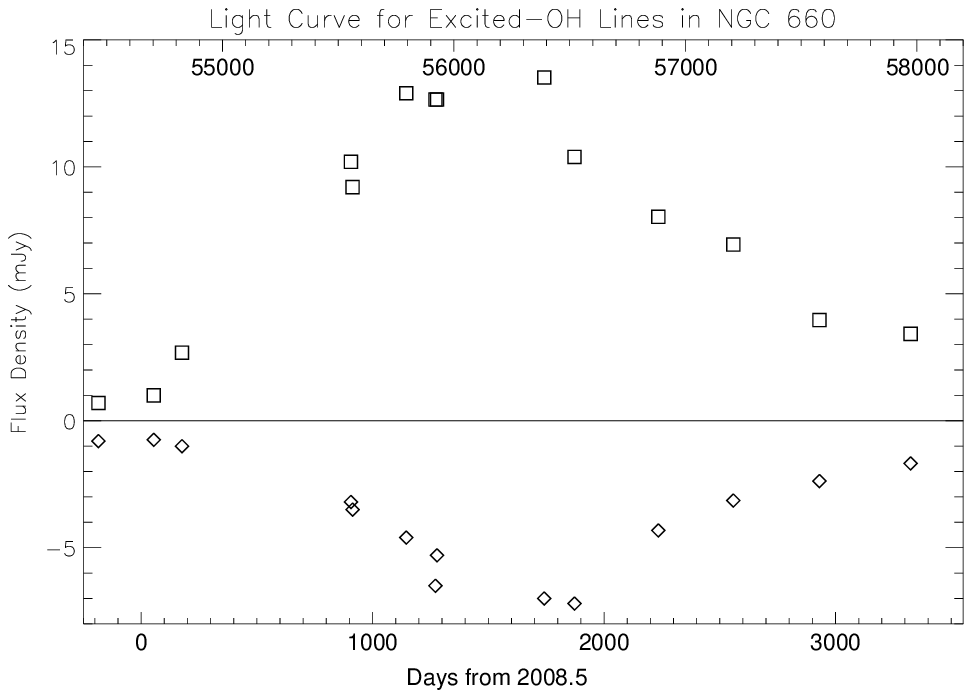}
\vspace{-0.25cm}
\caption{
(Left) The 4.7-GHz continuum light-curve of the NCC.  (Right) The light-curves of the excited-OH lines. Here, the data at the top (squares) represent the mean peak intensities of the strongest emission components at 4750 and 4765~MHz, while the symbols at the bottom (diamonds) show the mean peak intensity of the two 4660-MHz absorption components. The lower horizontal axis on both plots shows days from 2008.5,  while the top axes show MJD.}
\label{fig2}
\end{figure*}

During the final third of 2011, all three transitions had
become even stronger.  Additionally, a further manifestation of unusual
activity was discovered from the 2011 observations. When the C-band
continuum flux densities for each of our epochs were computed, a
steady increase over the 4-yr period from 2007 was seen.  Fig.
\ref{fig2} (left) shows the 4.7-GHz flux density of this ``new
continuum component'', (NCC), derived by subtracting the pre-outburst
spectrum (using a best fit to all published pre-2008 flux density
values), from our measurements. Fig.~\ref{fig2} (right) displays the
intensity changes of the excited-OH lines at a heliocentric 
velocity resolution of 10~km\,s$^{-1}$ via the mean peak intensity of the strongest pair of
emission components in the 4750- and 4765-MHz lines, and the mean of the
peak absorption intensities for the two absorption components seen at
4660~MHz. 

\begin{figure}
\plotone{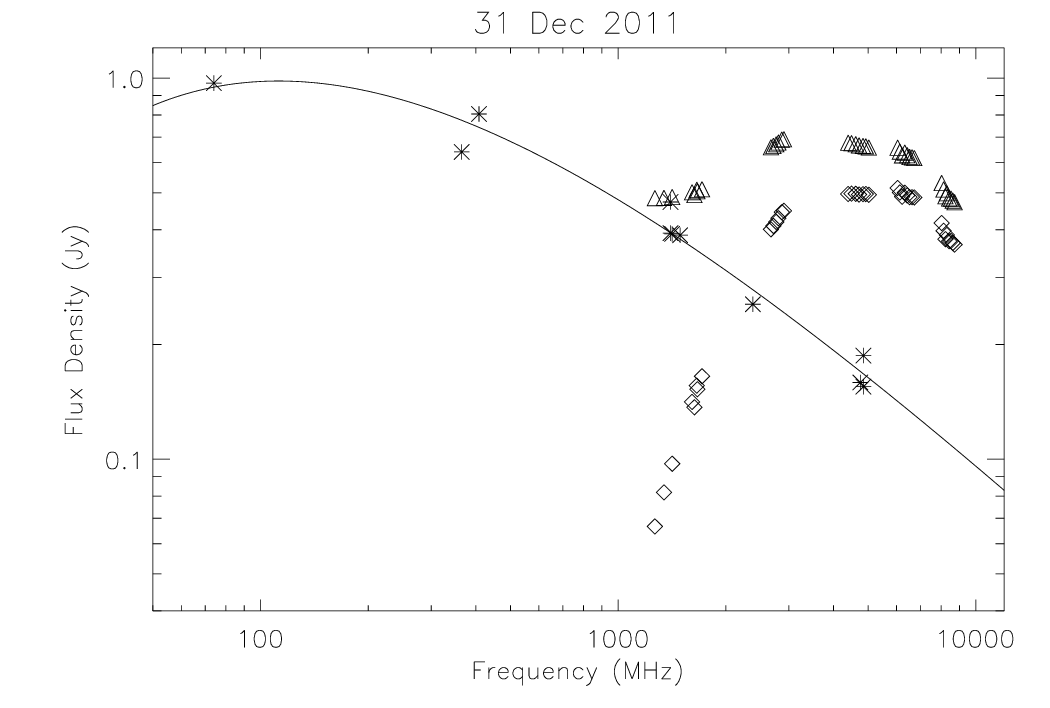}
\plotone{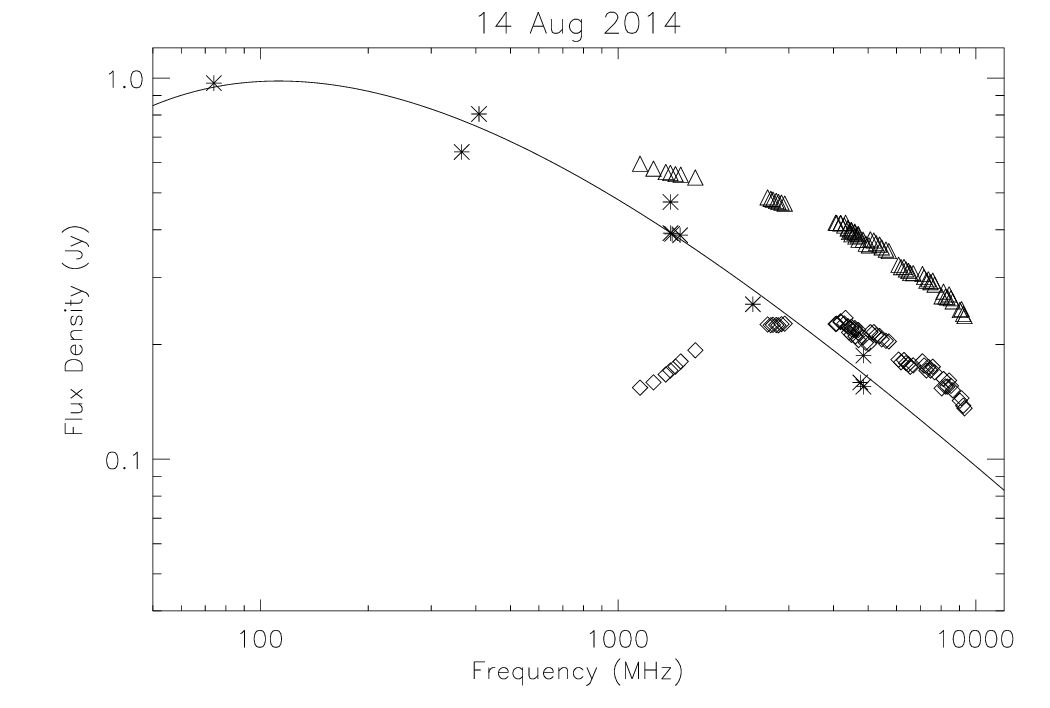}
\caption{The radio continuum spectrum of NGC~660 on 31 Dec 2011, near the  
epoch of the peak of the outburst (Top), and on 14 Aug 2014 (Bottom). Stars denote pre-outburst flux densities from the literature, with the solid line showing a spectral fit to this data. Triangles mark total measured flux densities, while diamonds show these total values minus the pre-outburst spectral fit, i.e. the spectrum of the transient New Continuum Component (NCC) alone.}
\label{fig3}
\end{figure}

The radio continuum emission of NGC~660 between
1.25 and 8.7~GHz obtained on 31 Dec 2011, again 
after subtracting the pre-outburst spectral fit to
obtain the detailed spectrum of the NCC is shown in Fig.~\ref{fig3} (Top), 
by diamond symbols. This reveals a GHz-Peaked Spectrum (GPS) form, peaking
close to 5~GHz and falling very rapidly towards lower frequencies. 
Fig.~\ref{fig3} (Bottom) presents the similarly derived
spectrum for the NCC made from data acquired on 14 August 2014. This
still shows a GPS form, although far less peaked than on the earlier
occasion. For 31 Dec 2011, we derive a spectral index (S $\propto \nu^{\alpha}$) for the NCC 
at frequencies below its peak emission of $\alpha^{2.76}\!\!\!\!\!\!\!\!_{1.42} \sim +2.2$,
consistent with total free-free or
synchrotron self absorption. By 14 Aug 2014, 
this had changed to $\alpha^{2.76}\!\!\!\!\!\!\!\!_{1.40} \sim +0.4$. At this later date,
the spectral index for frequencies above the emission peak of the NCC was
$\alpha^{9.62}\!\!\!\!\!\!\!\!_{5.03} \sim -0.63$, characteristic of optically-thin 
synchrotron emission. The NCC emission peak on 31 Dec 2011 was broadly situated at 
$5.5 \pm 1.0$ GHz, but had moved down to $4.0 \pm 1.0$ GHz by 14 Aug 2014. We note that such behavior is expected for a cloud of synchrotron-radiating relativistic particles that is initially opaque out to high radio frequencies, but which with expansion becomes optically-thin at successively lower frequencies \citep{vdl66}. 

\begin{figure}
\plotone{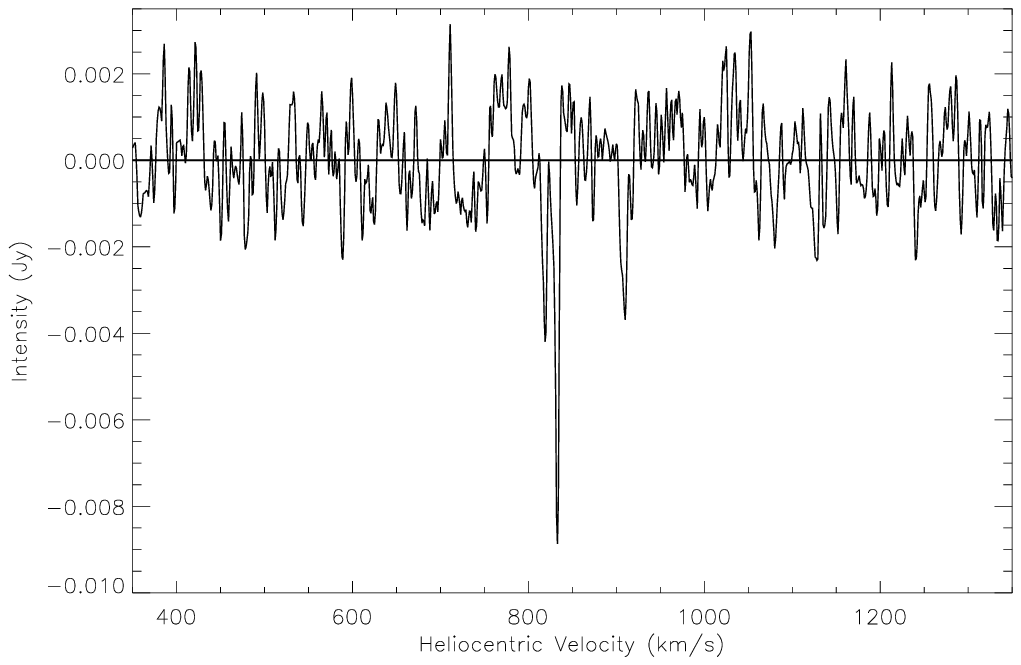}
\caption{The spectrum of H$_2$CO against the NCC. This was obtained by 
subtracting the line spectrum of Dec 2007 (i.e. pre-outburst) from the 
spectrum of Dec 2011, and shows new sharp absoption features.}
\label{fig4}
\end{figure}

While our pre-outburst 4.8-GHz H$_{2}$CO absorption spectrum agrees well with published data, the post-outburst spectrum shows detailed differences.  Subtracting the pre- and post-outburst spectra for this transition reveals 3 narrow absorption lines against the continuum
emission of the NCC, (Fig.~\ref{fig4}).  In addition, in late-Dec 2011 we obtained a small number of L-band spectra for NGC~660. For HI, a narrow absorption feature is clearly seen at the velocity of the deepest new H$_{2}$CO absorption line seen in Fig.~\ref{fig4}. The OH mainlines at 1665 and 1667~MHz show somewhat different detailed profiles to those previously published by \cite{ara04}.  Much more remarkably, the 1612- and 1720~MHz satellite OH lines are not only stronger than the main lines (up to 27~mJy) and show multiple components, but are closely conjugate, (i.e.  emission at 1612 MHz shows absorption of identical line profile at 1720~MHz, and vice versa). The physical explanation for their ``conjugate" nature depends upon how the two higher (hyperfine) energy levels of the OH molecule's ground state, i.e. $^2\Pi_{3/2} + F=2$, and $F=1$ are populated. The two possible pathways are, (1) intra-ladder $^{2}\Pi_{5/2} \rightarrow ^{2}\Pi_{3/2}$ at 119\,$\mu$m and the cross-ladder $^{2}\Pi_{1/2} \rightarrow ^{2}\Pi_{3/2}$ at 79\,$\mu$m. Excited OH molecules will cascade down to the ground level via either of these two transitions. With the applications of the selection rules (that parity has to change and $|\Delta F|  = 1$ or $0 $), the intra-ladder cascade would then cause overpopulation of the $F=2$ level, resulting in the 1720-MHz line being observed in emission and the 1612-MHz line in absorption. In the case of cross-ladder pumping, the situation will be reversed with the $F=1$ level being overpopulated leading to the 1612-MHz line appearing in emission and 1720-MHz line in absorption. Which route is more relevant depends upon the details of the pumping \citep{elitz1992}.
More details concerning the physical process leading to conjugate satellite transitions are 
discussed by \citet{lan95} and in references therein. Examples of such lines 
in Galactic interstellar clouds can be found in \cite{goss1968} and \cite{cashay75}, 
and for extragalactic sources such as Cen\,A \citep{lan95} and PKS~1413+135 \citep{kanek2004}.

The presence of such effects can be used to determine constraints on the OH-column 
density in the emitting/absorbing region. We cannot say whether these conjugate satellite 
OH lines have appeared since the occurrence of the outburst; certainly their velocities differ
from those of both the new excited-OH components, and the H$_{2}$CO absorption lines seen against the NCC. The full spectral-line data for the source as a function of observing epoch will be presented in a
future publication.

Preliminary results from the observations were presented by \citet{2013AAS...22115706M}. Following that announcement, 
observations made with e-MERLIN and the European  VLBI Network by \citet{2015MNRAS.452.1081A} confirmed that the outburst was a nuclear event. However, these lacked the spatial and spectral resolution of our HSA observations, and the time-series of our Arecibo observations, while concluding that the outburst
was likely to be the onset of a new period of AGN activity.

\section{Interferometric Follow-up}

Two scenarios that could be employed  as explanations
for the outburst in NGC~660 are:
\begin{itemize}
     \item Firstly, that this represents a heavily-obscured supernova (SN) outburst occurring in 2007--2008. Adopting the Tully-Fisher $z$-independent distance for NGC~660 of $14.7 \pm 2.2$~Mpc \citep{spr09} gives a 5-GHz continuum luminosity for the NCC of $1.3 \pm 0.2 \times 10^{22}$~W/Hz. However,  
     \citet{bie21} use their 2-10~GHz sample of 294 core-collapse SNe to derive a median spectral luminosity at peak of 10$^{18.5\pm1.6}$~W/Hz, with SNe of class~IIn reaching peak luminosities of 10$^{19.5\pm1.1}$~W/Hz. The NCC would thus represent an unusually luminous core-collapse radio SN. To date, only a single radio detection has been made of a Type-1a~SN \citep{Kool23, Yang23}. This was classified by \citet{Kool23} as being of Type~1a-CSM, developing an atypical Type~1a optical spectrum as it evolved. Both \citet{Kool23} and \citet{Yang23} find a 5.1-GHz continuum luminosity of $\sim 1.5 \times 10^{20}$~W/Hz. Were the  NCC to be an SNR that had been expanding freely at the canonical value of $10^4$~km/s, \citep [see] [who measured an expansion velocity of $9850 \pm 1500$~km/s for a young SN in the starburst galaxy M82] {Ped99}, for 1000 days, it would then have presented a diameter of $\sim$1~milliarcsec (mas) and would have been expected to show a ring or disk structure. However, this would have been barely resolved by terrestrial VLBI 
     at cm-wavelengths such as we present in Section~\ref{HSA-VLBI} and would have appeared as a quasi-compact
     source containing essentially the total flux density of the NCC. We detect no such component in our VLBI images of Subsection~4.2.
      \item The second possible explanation is that we are seeing the result of an outburst
originating at the very center of NGC~660 that commenced 3-4 yr
before the 2011/2012 epoch. For this, a core-jet appearance would be expected, with an
expansion velocity higher than for the SN case.  With  the light-curves
of the NCC and the OH features being so similar, the gas
involved in the molecular-line emission must be no more than a light month  or so
separated from the location of the continuum emission. Hence, the OH emission would be
expected to be distributed over a broadly similar area to the continuum
feature.
\end{itemize}

\subsection{The VLA Observations}

Given these scenarios, we decided to use VLBI imaging to help determine what
was actually happening within NGC~660.  To obtain a sub-arcsecond
position for the NCC, such as was necessary for the high-resolution VLBI
follow-up, we were first granted observing time with the VLA of the NRAO, Socorro, NM.

The VLA observations were carried out in a single 1.5-hr session on 2012
May 6 in the DnC configuration. Two 1-GHz wide bandwidths were observed
simultaneously, centered at 8.5 and 11.5~GHz. The WIDAR correlator was
configured to deliver $8 \times 128$~MHz sub-bands in each 1-GHz window for all
four polarization products (RR, LL, RL, LR). Each 128-MHz sub-band was
further sub-divided into 64 spectral channels. The standard source
3C48 was used to calibrate the absolute flux density scale, with
the source J0121$+$1149 being used as the complex gain calibrator.  The
total on-source time was about 55~minutes. At the time of the
observations, ten of the antennas were still equipped with the legacy
X-band receivers, which were sensitive only in the 8$-$8.8~GHz
frequency range, while the remaining antennas were equipped with the
new wide-band X-band receivers that cover the frequency range
8$-$12~GHz. Thus, the data for 8$-$8.8~GHz were from all the antennas
that participated in the observations, while the data for 8.8$-$9~GHz
and 11$-$12~GHz were only from the antennas that had the new receivers.

Editing, calibration, imaging and deconvolution of the data were
performed using the Astronomical Image Processing System (AIPS) of 
the NRAO \citep{2003ASSL..285..109G}.
After applying the phase and amplitude gain corrections from
J0121$+$1149 to the NGC~660 data, the target data were split into two files
corresponding to the $2 \times 1$-GHz frequency windows, and averaged in
frequency. Both phase and amplitude self-calibration, plus imaging,
were performed in an iterative cycle for each separate 1-GHz window to 
further improve the image quality.

\begin{figure*}
\plottwo{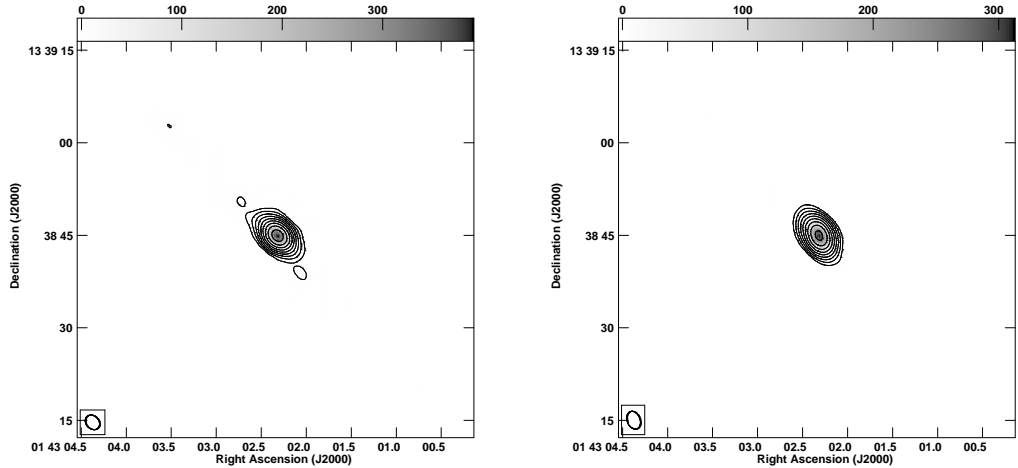}{Fig5b.eps}
\caption{Continuum images of NGC~660 at 8.5~GHz (Left) and 11.5~GHz (Right). 
For the 8.5-GHz image, the rms noise level is 75~$\mu$Jy~beam$^{-1}$ and the 
restoring beam size is $2\farcs59 \times 2\farcs06$  (P.A.$= 43\arcdeg$). For 
the 11.5-GHz image, the rms noise level is 105~$\mu$Jy~beam$^{-1}$ and the 
restoring beam size is $3\farcs00 \times 2\farcs08$  (P.A.$= 22\arcdeg$).  
The contour levels are at $-$1, 1, 2, 4, ..., 256~mJy~beam$^{-1}$. 
The gray-scale range is indicated at the top of each image in units of 
mJy~beam$^{-1}$.}
\label{fig5}
\end{figure*}

Figure \ref{fig5} shows the final continuum images of NGC~660 at 8.5
and 11.5~GHz. The rms noises in these images are 75 and
105~$\mu$Jy~beam$^{-1}$, with restoring beams of $2\farcs59 \times
2\farcs06$ (P.A.$= 43\arcdeg$) and $3\farcs00 \times 2\farcs08$ (P.A.
$= 22\arcdeg$), respectively. A two-dimensional Gaussian function was
fitted on the continuum source in each image. The best-fit
positions of the source are $\alpha=01^{\rm h}43^{\rm m}02\fs320$,
$\delta=13\arcdeg38\arcmin44\farcs96$ and $\alpha=01^{\rm h}43^{\rm
m}02\fs319$, $\delta=13\arcdeg38\arcmin44\farcs97$ in the 8.5- and
11.5-GHz images respectively. These positions agree to better than 0.1~arcsec
with that measured for the pre-outburst radio core at 8.4~GHz from VLA A-array data by 
\citet{fil02}. As the majority of the flux density in our images will have 
originated from  the NCC, this confirms its location at the core of NGC~660.

At 8.5~GHz, the peak and total flux densities were $385.8 \pm
0.1$~mJy~beam$^{-1}$ and $414.0 \pm 0.2$~mJy, respectively. The source
was resolved at this frequency, with a nominal deconvolved size of
$0\farcs89 \times 0\farcs35$ (P.A. $= 49\arcdeg$). At 11.5~GHz, the
peak and total flux densities were $309.1 \pm 0.1$~mJy~beam$^{-1}$ and
$333.6 \pm 0.2$~mJy, respectively. The source was resolved at this
frequency, with a nominal deconvolved size of $0\farcs96 \times
0\farcs42$ (P.A. $= 48\arcdeg$).

\subsection{The HSA (VLBI) Observations}
\label{HSA-VLBI}

In July 2012, High Sensitivity Array (HSA) VLBI observations  at C- and X- band 
were made to image 
the NCC continuum structure with $\sim$1-mas resolution, and to investigate 
the excited-OH emission/absorption and H$_2$CO
absorption against the NCC.   These observations used the HSA with stations from
the Very Long Baseline Array (VLBA), along with the 100-m Effelsberg radio 
telescope, the 305-m Arecibo telescope and the 100-m Green Bank Telescope. For the C-band observations, the
Saint Croix and Los Alamos VLBA stations were not available, giving a total 
of 11 antennas in the array. For the X-band observations all ten VLBA stations
participated.

\begin{figure}[b]
\plottwo{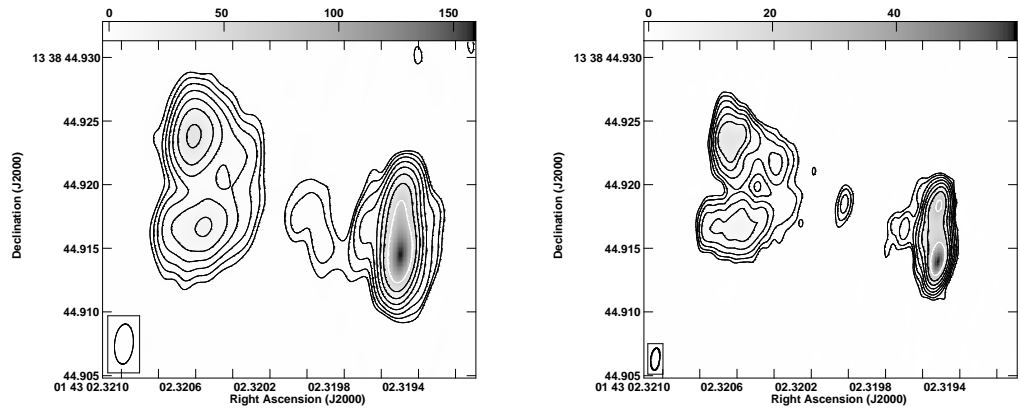}{VLBA_X_edited.eps}
\caption{\footnotesize
(Left) The HSA continuum images of NGC~660 at 4.7~GHz, HPBW = 3.1 $\times$  1.5
milliarcsec at P.A. = --4\fdg9, and (right) at 8.4~GHz, HPBW =
1.6 $\times$  0.7 milliarcsec at P.A. =  --5\fdg9.
Contour levels are at 2.5$\sigma \times$ 2.5, 5, 10, 20, 40, 80, 160, 320 with $\sigma$ = 80 $\mu$Jy at 4.7~GHz and $\sigma$ = 36 $\mu$Jy at 8.4~GHz. The gray-scale range is indicated at the top of each image in units of 
mJy~beam$^{-1}$.}
\label{fig6}
\end{figure}

Analysis of the C- and X-band data has produced remarkable
continuum images, (Fig.~\ref{fig6}).  The milliarcsec structure shows a triple
source consisting of a  flat-spectrum, ($\alpha \sim 0.0$),  
compact central component, with a
pair of more diffuse, steep-spectrum 
features disposed roughly E-W on
either side.  The western feature consists of a bright pair of ``hot-spots'', 
with no jet emission being seen. In contrast, the
eastern feature has the form of an edge-brightened jet, 
with the center lines of the edge-brightened
features pointing through the central source, towards the opposing
western hot-spots.  The overall emission distribution seems to be consistent with the canonical
structure of a classical radio source \citep{con16}, perhaps in this case with
a pair of rapidly-precessing, relativistic jets
emanating from the active nucleus, with the jet directions precessing along the surfaces of 
diametrically-opposed cones. 
The jet to the east would be the approaching, Doppler-boosted jet, with 
emission from lines of sight passing through the edges of 
that cone being further enhanced by the greater geometrical, line-of-sight depths 
encountered there. In contrast, the receding jet to the west would be Doppler-diminished 
and hence not seen. However, 
this western jet would be responsible for the western hot-spots where it
interacts with the circumnuclear medium.

As we present only single-epoch VLBI observations in Figure 6, it may
be asked whether the core, jet and hotspot emission revealed
there might not pre-exist the present outburst? However, two pieces of
evidence suggest this not to be the case.
Firstly, \citet{filho04} observed the core of NGC 660 with the VLBA at
C-band in September 2001, setting a $5\sigma$ upper limit to the peak
flux of 0.5 mJy/beam and not making any detection. Their observations
should have easily detected (with SNR $>$ 70) the hotspots seen on
both sides of the core were these to have been present at that time. 
Secondly, linear interpolation between the two nearest Arecibo single-dish 4.7-GHz
continuum flux density estimates for the NCC yields a value of 447~mJy for
mid-July 2013. The total flux density integrated from our 4.7-GHz VLBI 
image of Figure 6 (left) at that epoch is 458~mJy. The outstanding
agreement of these two values indicates the identity of the Arecibo NCC
and the VLBI structure revealed by our HSA observations.

\begin{figure}[ht]
\plotone{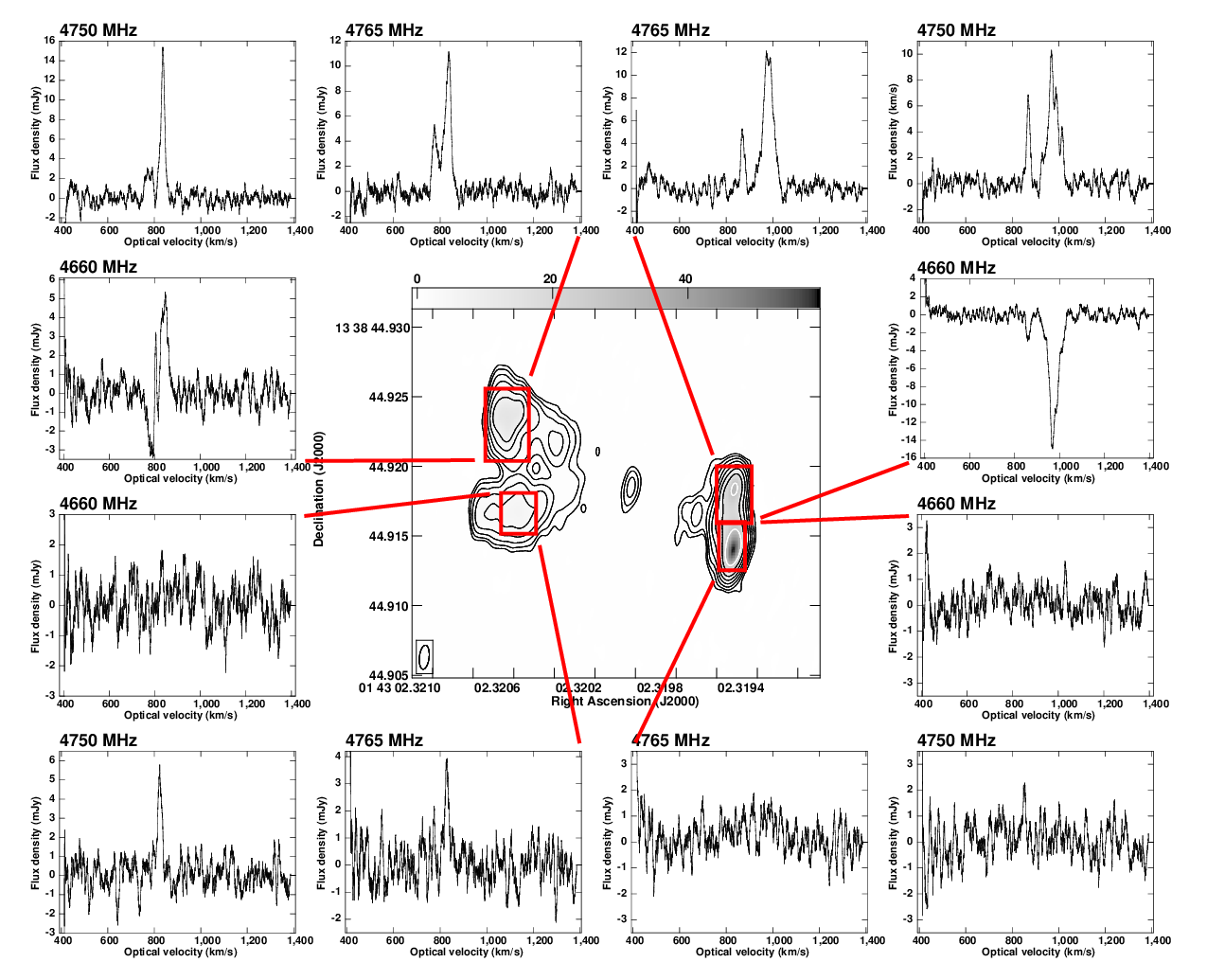}
\caption{The integrated excited-OH lines from the regions of the four
``hotspots'' seen in the X-band image of Fig.~\ref{fig6}. 
The higher velocity components occur exclusively against the north-western hotspot,
while those at lower velocities are mainly found against the two eastern
hotspots.}
\label{fig7}
\end{figure}

The July 2012 C-band VLBI observations showed that the excited-OH and and 
H$_{2}$CO lines found at Arecibo were coincident with the continuum
hotspots of the NCC
revealed by the HSA observations, confirming at least near physical coincidence between these
continuum features and the spectral outburst in NGC660 (Fig.~\ref{fig7}).
The velocity structure of the OH lines from the C-band data cubes shows that those associated with
the eastern hotspots lie at considerably lower radial velocities than those associated with 
the region of the north-western hotspot. This is consistent with the above 
suggestion that the eastern VLBI components represent the approaching jet, with the western VLBI hotspots 
defining the receding jet. 
A full analysis of the spectral properties will be presented in a later paper.

Assuming the central compact component seen at X-band to be the galaxy's active nucleus, 
the angular distances between it and the peaks in the four ``hotspots'', 
as measured from the 8.4-GHz image of Fig.~\ref{fig6}, are:
\begin{itemize}
\item NE hotspot $\leftrightarrow$ central source = 10.2~mas (11.4~mas to the outer edge)
\item SE hotspot $\leftrightarrow$ central source = 10.5~mas (11.6~mas)
\item NW hotspot $\leftrightarrow$ central source = 7.4~mas (8.4~mas)
\item SW hotspot $\leftrightarrow$ central source = 8.4~mas (9.1~mas)
\end{itemize}

The opening angle from the central component to the two hotspots on either 
side are $\approx 34^{\circ}$ to the eastern hotspot pair, and 
$\approx 23^{\circ}$ to the western pair.

\subsection{Implications of the VLBI Results}

If the NCC is an intrinsically symmetric double radio source, in which the outer
components advance from the nucleus at a velocity of $v = \beta c$ along an axis inclined
at an angle $\phi$ to the line of sight, then \citep{sai81}:
\begin{equation}
\theta_1 /\theta_2 = (1 + \beta \cos\,\phi)/(1 - \beta \cos\,\phi)
\end{equation}
where $\theta_1$ and $\theta_2$ are the apparent separations of the approaching and
receding components from the radio core. For the above displacements from the central source
(assumed to be the active nucleus), we have a ratio of 10.2/8.4 = 1.214 for the NE/SW hotspot pair and 
a ratio of 10.5/7.4 =1.419 for the NW/SE hotspot pair. These give $\beta \cos\,\phi = 0.097$ 
for the first pair and $\beta \cos\,\phi = 0.173$ for the second pair. Combining these gives 
a value of:

 $\beta \cos\,\phi = 0.135 \pm 0.038$. 

In addition, for the Tully-Fisher distance of 13.6 Mpc, the projected distance 
from the core to the approaching hotspots is $0.68\pm 0.01$ pc. If the approaching 
jet was launched from close to the core, and had been flowing out for T yr at the
epoch of the HSA observations, then its apparent velocity in the plane of the sky 
would be $6.65 \pm 0.10 \times 10^5 / \rm{T}$ km\,s$^{-1}$, giving  $1.6 \pm 0.2 
\times 10^5$ km/sec for T=$4.1 \pm 0.5$ yr, (assuming an epoch of $2008.5 \pm 0.5$, 
i.e. midway between the December 2007 and December 2008 observations, 
for the outburst date, see Fig.~\ref{fig1}). 
This gives the apparent perpendicular velocity as a fraction of 
the speed of light, $\beta_{\perp} = 0.53 \pm 0.07$.
 
For a component moving at an angle $\phi$ to the line of sight  \citep{con16}, 
$\beta_{\perp}$ is given by: 
\begin{equation}
\beta_{\perp} = \beta \sin\,\phi /(1 - \beta \cos\,\phi)
\end{equation}
Substituting $\beta \cos\,\phi = 0.135 \pm 0.038$ and $\beta_{\perp} = 0.53 \pm 0.07$ into this 
equation yields $\beta \sin\,\phi = 0.46 \pm 0.14$.

The above values of $\beta \cos\,\phi$ and $\beta sin\,\phi$ yield 
$\tan\,\phi = 3.4\pm 1.2$ and thus $\phi = 74^{+4}_{-8} $ degrees and 
$\beta = 0.48 \pm 0.16$. Of course, how much faith can
be placed on these values is dependent on many assumptions:
Were the jets launched from the proximity of the galaxy nucleus, and when?
Are the opposing jets co-linear? Is the ambient medium symmetric on
opposite sides of the nucleus? However, it is suggestive that the new
continuum component is mildly relativistic, and moving closer to the
plane of the sky than to the line of sight.

The mean brightness of the jet towards the east is at least 4 times
greater than the jet towards the west, possibly much more. It is of
interest \citep{kel88} that a jet with a positive velocity
component towards the observer is Doppler boosted by a factor of
$[\gamma^{(\alpha - 3)} (1-\beta cos(\phi))]^{(\alpha - 3)}$, where
$\gamma = (1-\beta^2)^{-0.5}$ and $\alpha$ is the spectral index
of the jet. The jet pointing away from the observer will
be diminished by a similar factor.  The value of $\alpha$ for the NCC
between 7.1 and 9.6 GHz  at that epoch was $\alpha \sim -0.9$ from our Arecibo monitoring of the
source flux density.  For $\phi \approx 74 \arcdeg$ and $\beta \approx
0.48$, the brightness ratio of the opposing jet features would be $\approx 25$.

\section{Discussion}

Given the consistency of the HSA continuum image with an outburst at
the nucleus of NGC~660, the existence of mildly relativistic jets, 
and the lack of a single quasi-compact component containing the majority of
the NCC flux density,
the hypothesis that the NGC~660 outburst
represents a heavily-obscured supernova explosion now seems untenable.
Indeed, we appear to be seeing an outburst in the nucleus of NGC~660
that commenced in the radio in 2008. The morphology revealed by our VLBI
observations is consistent with a rapidly-precessing, diametrically-opposed 
pair of radio jets from the
core of the galaxy. However, the question remains: as to what caused these jets? Plausible
explanations for the NGC~660
outburst might be that the observed VLBI jets have been launched by either (a)
the infall of a gas cloud on to the central super-massive black hole (SMBH),
or (b) a ``tidal disruption event" (TDE) caused by a star passing
within the ``tidal radius'' of the SMBH and being torn asunder, with
part of its material being ingested by the BH, and the rest being
ejected as a high velocity jet \citep{ree88, zau11, bow13}.
If this were a TDE, then the evolution of the source morphology
and the detailed line and continuum radio light-curves would provide detailed
tests for many of the models proposed for these objects.

The possibility of the disruption of a very massive star is non-negligible given the
presence of a nuclear starburst in this galaxy. In addition, the
presence of a massive polar ring around the main galaxy disc indicates
a collision with a massive companion within the past few billion years
\citep{van95}. Indeed, van Driel et al. note that the shape
of NGC660's inner bulge may indicate peculiar stellar orbits in the
bulge, possibly making TDEs not unlikely events in this galaxy.

The major challenge for explaining the NGC~660 outburst as a TDE is that
most such events, at all wavelength ranges, have evolved on timescales
of less than one year \citep{ale20} and have light-curves with a rapid initial
rise followed by a slow decline. Most TDEs have also been found in quiescent,
post-starburst galaxies \citep{fre17}. 
In contrast, the NGC~660 outburst occurred in a star-forming LIRG, peaked
three years after initial detection (four years after the previous non-detection)
and was followed by us for almost a decade. While a likely TDE
lasting for over a decade was discovered in the soft X-ray
range within a dwarf starburst galaxy by \citet{lin17}, which the authors 
suggest represented the disruption of a very massive star, this event had a fast 
rise time in luminosity ($<$ 4 months), and then underwent a very slow decay that 
was followed for over 10 years, very different in shape to the light-curve seen 
in NGC~660 (Fig.~\ref{fig2}). There have also been observations of late
radio-emission from TDEs \citep[e.g.][]{2021NatAs...5..491H,2021ApJ...920L...5H,2022ApJ...938...28C}, with continuum 
peaks as late after optical detection as the time from first (radio) detection 
to the peak in NGC~660. However, in these cases there was no detectable radio
emission until close to the time of the radio peak, followed by a steep rise
ascribed to the late launching of a jet. This scenario does not fit with what has
been seen in NGC~660. Further,
as noted by \citet{2015MNRAS.452.1081A}, archival X-ray data show no sign
of a peak in X-ray luminosity as would be expected if this were a TDE.

The shape of the light-curve for this event points toward it being the accretion of a 
gas cloud rather than a TDE. This does not seem to be the dawn of a new period of 
sustained AGN activity, as suggested by \citet{2015MNRAS.452.1081A}, as the 
continuum light curve peaked in late 2011 and fell back to below 25\% of that peak
by 2017, while the OH emission and absorption peaked by the end of 2012 and have also
fallen back to below 25\% of those peaks. Rather, it seems likely that this outburst
was a single, extended event -- most likely the ingestion of a gas cloud over a
prolonged period.

\section{Conclusion}
We serendipitously detected a quasi-simultaneous continuum and spectral-line outburst in NGC~660 during
a line survey with the 305-m Arecibo radio telescope. Over a decade, we have monitored both 
the continuum outburst, and the variable emission and absorption spectra of 
the excited-OH molecule at C-band. High resolution VLBI follow-up has 
revealed these remarkable transient features to be located at the center of the galaxy and to have a morphology 
consistent with a rapidly-precessing, two-sided jet. Analysis of the approaching and receding
components of the jet suggests that it is closer to the plane of the sky than to the line of 
sight and that it is mildly relativistic. The source of the outburst is likely to be the accretion
of either a gas cloud or debris from a 
tidal disruption event onto the central supermassive black hole. From the shape of the light-curve, accretion of a gas cloud
would appear the more likely of these two possibilities.

\section{Acknowledgement}
We thank an anonymous referee for a number of useful suggestions which have
considerably improved the paper.
Arecibo Observatory was operated
by Cornell University until 2011 under a cooperative agreement with
the National Science Foundation (NSF). From 2011 until 2018 by SRI
International under a similar cooperative agreement, in alliance with
Ana G.  M\'{e}ndez-UMet and USRA.  From April 2018 until August 2023 it was operated by the
University of Central Florida under a cooperative agreement with the NSF
and in alliance with Universidad Ana G. M\'{e}ndez and Yang Enterprises,
Inc.

The National Radio Astronomy Observatory
(NRAO) is a facility of the NSF operated under cooperative agreement by
Associated Universities, Inc.

\end{document}